\documentclass[aip,jcp,reprint,amsmath,amssymb,floatfix,citeautoscript,nofootinbib]{revtex4-2}
\usepackage{cancel}
\usepackage{amsmath}
\usepackage{physics}
\usepackage{graphicx}
\usepackage{amssymb}
\usepackage[english]{babel}
\usepackage{amsthm}
\usepackage{bm}
\usepackage{booktabs}
\usepackage{threeparttable}
\usepackage{dcolumn}
\newcommand{\ttext}[1]{\multicolumn{1}{c}{#1}}
\newcolumntype{x}[1]{D{.}{.}{#1}}
\usepackage{ragged2e}
\usepackage{txfonts}
\allowdisplaybreaks
\usepackage{mathtools}
\usepackage{color}
\usepackage[usenames,dvipsnames]{xcolor}
\definecolor{myblue}{rgb}{0,0,1}
\usepackage[breaklinks=true,colorlinks=true,linkcolor=myblue,urlcolor=myblue,citecolor=myblue]{hyperref}
\DeclarePairedDelimiterX\pint[2]{(}{)}{#1 \delimsize\vert #2}

\newcommand{\vk}{{\bm{k}}}
\newcommand{\vG}{{\bm{G}}}

\begin{document}

\title{
Can spin-component scaled MP2 achieve kJ/mol accuracy for cohesive energies of molecular crystals?
}
\author{Yu Hsuan Liang}
\affiliation{Department of Chemistry, Columbia University, New York, NY 10027 USA}
\author{Hong-Zhou Ye}
\email{hzyechem@gmail.com}
\affiliation{Department of Chemistry, Columbia University, New York, NY 10027 USA}
\author{Timothy C. Berkelbach}
\email{t.berkelbach@columbia.edu}
\affiliation{Department of Chemistry, Columbia University, New York, NY 10027 USA}

\begin{abstract}
Achieving kJ/mol accuracy in the cohesive energy of molecular crystals, as necessary for crystal structure prediction and 
the resolution of polymorphism, is an ongoing challenge in computational materials science.
Here, we evaluate the performance of second-order M{\o}ller-Plesset perturbation theory (MP2), including its spin-component scaled
models, by calculating the cohesive energies of the 23 molecular crystals contained in the X23 dataset.
Our calculations are performed with periodic boundary conditions and Brillouin zone sampling,
and we converge results to the thermodynamic limit and the complete basis set limit to an accuracy of about 1~kJ/mol (0.25~kcal/mol),
which is rarely achieved in previous MP2 calculations of molecular crystals.
Comparing to experimental cohesive energies, we find that MP2 has a mean absolute error of 12.9 kJ/mol, which is comparable to that of DFT using the PBE functional and TS dispersion correction.
Separate scaling of the opposite-spin and same-spin components of the correlation energy, with parameters previously determined for molecular interactions, reduces the mean absolute error to 9.5~kJ/mol,
and reoptimizing the spin-component scaling parameters for the X23 set further reduces the mean absolute error to 7.5~kJ/mol.

\end{abstract}

\maketitle

Molecular crystals are periodic arrangements of molecules bound by weak, noncovalent interactions.
They play an important role in the pharmaceutical industry, and their unique electronic and optical properties are of interest for the development of molecular optoelectronics~\cite{beran_modeling_2016}.
The small energy scale of their interactions gives rise to polymorphism, where one molecule crystallizes into more than one structure, which have different properties such as their stability or solubility~\cite{beran_modeling_2016,marom_many-body_2013}.
The prediction of relative stabilities of competing crystal structures is a computational grand challenge due to the chemical nature and magnitude of the competing molecular interactions,
and a successful methodology would have major impact on basic and applied research.

Density functional theory (DFT)~\cite{kohn_self-consistent_1965,hohenberg_inhomogeneous_1964} is among the most popular methods for predicting material properties from first-principles due to its low cost and widely available implementation in popular software packages.
However, DFT with the commonly employed semi-local~\cite{kristyan_can_1994} or hybrid functionals~\cite{kozuch_dsd-pbep86_2011,heyd_hybrid_2003,broqvist_hybrid-functional_2009} does not accurately capture non-covalent interactions such as dispersion, which are essential for molecular crystals.
Efforts to address this deficiency include the empirical `-D' corrections of Grimme~\cite{grimme_accurate_2004}, the exchange-hole dipole moment (XDM) method of Becke and Johnson~\cite{becke_exchange-hole_2005}, and the van der Waals $C_6$ model of Tkatchenko and Scheffler~\cite{tkatchenko_accurate_2009}.

An alternative to DFT is a wave-function based approach, which naturally includes dispersion interactions at the post-Hartree-Fock (HF) level and which is, in principle, systematically improvable.
Among such methods, second-order M{\o}ller-Plesset perturbation theory (MP2)~\cite{moller_note_1934} is the simplest correlation method
and its accuracy can be significantly improved by the separate scaling of the spin components of its correlation energy
(at least, for molecules)~\cite{grimme_improved_2003,grimme_accurate_2004,jung_scaled_2004,distasio_jr_optimized_2007}.
Despite the increasing use of periodic MP2 due to its relatively low cost, there are few systematic reports on its performance~\cite{maschio_periodic_2010,maschio_intermolecular_2011,del_ben_second-order_2012,cutini_assessment_2016,thomas_accurate_2018}, especially in the computationally demanding complete basis set (CBS)~\cite{shepherd_convergence_2012,booth_plane_2016,callahan_dynamical_2021,lee_approaching_2021,ye_correlation-consistent_2022,marsman_second-order_2009} limit and thermodynamic limit (TDL)~\cite{gruber_applying_2018,neufeld_ground-state_2022,gruneis_second-order_2010}.
Recently, our group reported such a study for semiconductors and insulators with strong covalent or ionic bonds~\cite{goldzak_accurate_2022}, finding that MP2 yields reasonable predictions for lattice constants, bulk moduli, and cohesive energies.
For example, the mean absolute error of the cohesive energy, compared to experimental values, was 22~kJ/mol; with separate scaling of the spin components, the error was reduced to 6~kJ/mol, which is better than that of good DFT functionals like PBEsol or SCAN.
In these strongly bound solids, the HF cohesive energy is qualitatively correct and constitutes about 60--70\% of the MP2 cohesive energy.
In contrast, in many weakly bound molecular crystals, the HF cohesive energy is only a small fraction of the total cohesive energy, providing a more challenging test of approximate theories of electron correlation.

Recently, our group reported a periodic MP2 study of the benzene crystal~\cite{bintrim_integral-direct_2022}, showing promising results and addressing potential basis set incompleteness and finite-size errors in literature MP2 values.
We found that MP2 overestimated the magnitude of the cohesive energy by about 18~kJ/mol and, again, that spin scaling could reduce the error to 3--5~kJ/mol.
Here, we extend this work and report tightly converged periodic MP2 calculations of the cohesive energies of the 23 molecular crystals contained in the X23 dataset
of Reilly and Tkatchenko~\cite{reilly_understanding_2013}, which built on the C21 dataset of Otero-de-la-Roza and Johnson~\cite{otero-de-la-roza_benchmark_2012}.

The crystals in the X23 dataset exhibit diverse bonding types, including dipole interactions, pure dispersion, and hydrogen bonding.
For select molecular crystals, coupled-cluster theory with single, double, and perturbative triple excitations [CCSD(T)] has achieved
kJ/mol accuracy in calculated cohesive energies~\cite{yang_ab-initio_2014,sherrill_2023}, but its high cost---scaling as $N^7$ with system size $N$---precludes routine application to systems with large unit cells.
Fifteen years ago, DiStasio and Head-Gordon targeted CCSD(T) accuracy with MP2 cost by optimizing the MP2 spin-scaling coefficients specifically for intermolecular interaction energies [SCS(MI)-MP2]~\cite{distasio_jr_optimized_2007} using the S22 dataset~\cite{jurevcka_benchmark_2006} of non-covalently bonded model complexes.
In the CBS limit, spin scaling reduced the mean absolute error with respect to CCSD(T) from 3.3~kJ/mol to less than 1~kJ/mol
and reduced the maximum error from 12~kJ/mol to 2~kJ/mol.
In this work, we test the transferability of this and other spin-scaling prescriptions for molecular crystals, whose many-body and long-range interactions may not be reflected in the dimers included in the S22 dataset.

Our calculations were performed using PySCF~\cite{sun_recent_2020,sun_pyscf_2018} with the all-electron cc-pV$X$Z ($X$=D,T,Q) basis sets~\cite{dunning_gaussian_1989}
and periodic Gaussian density fitting of electron repulsion
integrals~\cite{sun_gaussian_2017,ye_fast_2021,ye_tight_2021} with corresponding JKFIT auxiliary basis sets.
Core electrons were kept frozen during MP2 calculations.
For a fixed $k$-point mesh containing $N_k$ points sampled uniformly from the Brillouin zone,
the MP2 correlation energy per unit cell is
$E^{(2)} = E^{(2)}_\mathrm{os} + E^{(2)}_\mathrm{ss}$, with the
opposite-spin and same-spin components,
\begin{subequations}
    \begin{equation}
        E_{\mathrm{os}}^{\mathrm{(2)}} = -\frac{1}{N_k^3} \sideset{}{'}\sum_{\vk_i\vk_a\vk_j\vk_b}\sum_{iajb} T^{a\vk_a,b\vk_b}_{i\vk_i,j\vk_j} \pint{i^{\vk_i}a^{\vk_a}}{j^{\vk_j}b^{\vk_b}}
        \label{eq:EMP2_os}
    \end{equation}
    \begin{equation}
        E_{\mathrm{ss}}^{\mathrm{(2)}} = -\frac{1}{N_k^3} \sideset{}{'}\sum_{\vk_i\vk_a\vk_j\vk_b}\sum_{iajb} [T^{a\vk_a,b\vk_b}_{i\vk_i,j\vk_j}-T^{b\vk_b,a\vk_a}_{i\vk_i,j\vk_j}]\pint{i^{\vk_i}a^{\vk_a}}{j^{\vk_j}b^{\vk_b}},
        \label{eq:EMP2_ss}
    \end{equation}
\end{subequations}
where
\begin{equation}
T^{a\vk_a,b\vk_b}_{i\vk_i,j\vk_j} = \frac{\pint{i^{\vk_1}a^{\vk_2}}{j^{\vk_3}b^{\vk_4}}^*}
    {\varepsilon_a^{\vk_a}-\varepsilon_i^{\vk_i}+\varepsilon_b^{\vk_b}-\varepsilon_j^{\vk_j}},
\end{equation}
$\pint{i^{\vk_1}a^{\vk_2}}{j^{\vk_3}b^{\vk_4}}$ are electron repulsion integrals, and $\varepsilon_i^{\vk_i}$ are HF orbital energies.
The primed summation indicates conservation of crystal momentum such that $\vk_i+\vk_j-\vk_a-\vk_b=\vG$, where $\vG$ is
a reciprocal lattice vector.

For each $N_k$, we have verified that the HF energy is converged at the QZ level
and we extrapolate the MP2 correlation energy.
Specifically, we perform a series of MP2 calculations with increasing $X$, and we extrapolate the correlation energy to the CBS limit using the two-point $X^{-3}$ form~\cite{paier_screened_2006,broqvist_hybrid-functional_2009,sundararaman_regularization_2013},
\begin{equation}
    E^{(2)}(N_k,\mathrm{CBS}) = \frac{X^3E^{(2)}(N_k,X\mathrm{Z}) - Y^3E^{(2)}(N_k,Y\mathrm{Z})}{X^3-Y^3},
    \label{cbs-ex}
\end{equation}
where $X=3,4$ (i.e., TZ and QZ).
The CBS HF and MP2 energies are then extrapolated to the TDL assuming finite-size errors that decay as $N_k^{-1}$,
\begin{equation}
E(\mathrm{TDL},\mathrm{CBS}) = \frac{N_{k,1}E(N_{k,1},\mathrm{CBS})-N_{k,2}E(N_{k,2},\mathrm{CBS})}
    {N_{k,1}-N_{k,2}}
    \label{tdl-ex}
\end{equation}
which is consistent with our use of a Madelung constant correction for the integrable divergence in the HF
exchange~\cite{paier_screened_2006,broqvist_hybrid-functional_2009,sundararaman_regularization_2013}.

\begin{figure}[t]
\centering
    \includegraphics[width=8cm]{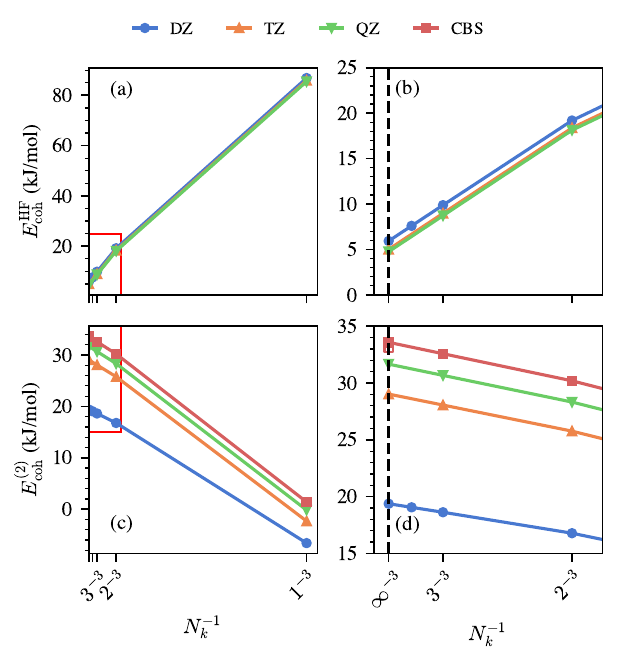}
\caption{Thermodynamic limit convergence of energies of the ammonia crystal using different basis sets.
HF cohesive energies and MP2 correlation contributions to the cohesive energies are shown in panels (a) and (c), respectively, with zoomed-in views shown in (b) and (d).
Hollow square in (d) indicates the estimate obtained from the composite correction scheme described in the text, which is accurate to about 0.5~kJ/mol.
}
\label{fig:ammonia_ex}
\end{figure}

The $k$-point meshes used for extrapolation are chosen commensurate with the shape of the unit cell.
For example, rhombohedral cells with lattice parameters $a=b=c$ suggest meshes like $N_k = 111, 222, 333$,
while hexagonal cells with $a=b>c$ suggests meshes like $N_k = 221, 332, 443$ (we write $N_k = abc$ as shorthand
for $a\times b\times c$).
Optimal $k$-point mesh pairs are determined by exploration of several appropriate $k$-point mesh pairs with a cheap minimal basis set.

An example of the CBS and TDL extrapolations is shown in figure~\ref{fig:ammonia_ex} for the ammonia crystal.
In this particular example, we chose $N_k=222, 333$ as the pair of $k$-point meshes for production calculations, as it provides a small error when compared to results obtained with larger meshes but smaller basis sets;
for example, in figure~\ref{fig:ammonia_ex}, we also present results with $N_k = 444$ in the DZ basis set,
showing that extrapolation with $N_k=222, 333$ is in good agreement with that from $N_k=333, 444$.

Large systems in the X23 dataset required special attention because their high computational costs precludes study of appropriately large basis sets and $k$-point meshes.
For example, the largest unit cell in the dataset, pyrazole, contains 80 atoms and 5872 basis functions at the QZ level.
Therefore, for large systems like these, we employed a composite correction scheme to estimate the TDL/CBS limit,
\begin{equation}
E(\mathrm{TDL},\mathrm{CBS}) \approx E(\mathrm{TDL}, \mathrm{TZ})
    + \left[ E(N_k, \mathrm{CBS}) - E(N_k, \mathrm{TZ})\right],
\end{equation}
the accuracy of which is illustrated in figure~\ref{fig:ammonia_ex} for ammonia and explored in more details for other crystals in the Supporting Information.
Moreover, for all calculations requiring more than 2000 GB of storage, we utilized the integral-direct algorithm implemented in PySCF~\cite{sun_pyscf_2018,bintrim_integral-direct_2022}.
In the Supporting Information, we provide further details on our TDL and CBS extrapolations, including the composite correction, which allow us to conclude that our final numbers are converged to about 2~kJ/mol.

Finally, we report the counterpoise corrected cohesive energy,
\begin{equation}    \label{eq:ecoh}
    -E_{\mathrm{coh}} = \frac{E_{\mathrm{cell}}}{N_{\mathrm{mol}}} - E_{\mathrm{mol+ghost}}^{\mathrm{crystal}}
        + \left(E_{\mathrm{mol}}^{\mathrm{crystal}} - E_{\mathrm{mol}}^{\mathrm{gas}}\right)
\end{equation}
where $E_{\mathrm{cell}}$ is the total crystal energy per cell, $E_{\mathrm{mol+ghost}}^{\mathrm{crystal}}$ is the energy of the molecule in its crystal geometry including basis functions from its first nearest-neighbor shell of ghost atoms, and the final term in parentheses (a molecular relaxation energy) is the energy difference between the molecule in the crystal geometry and its most stable gas phase geometry (without any ghost atoms).
We note our sign convention is such that the cohesive energy $E_\mathrm{coh}$ is the (positive) energy required to dissociate the crystal into its constituent molecules.

All geometries used are obtained from the original X23 paper~\cite{reilly_understanding_2013}, where all geometries (including cell parameters for crystals) were optimized with DFT
using the PBE functional and the TS dispersion correction (PBE-TS)~\cite{tkatchenko_accurate_2009}. Importantly, our own testing (see below) indicates that the use of alternative geometries, such as those taken directly from X-ray diffraction, can change the cohesive energy by 5~kJ/mol or more.
We compare our calculated cohesive energies to the revised X23 reference values~\cite{dolgonos_revised_2019}, which were obtained by correcting experimental sublimation enthalpies for temperature-dependent vibrational contributions, including thermal expansion.
 \begin{table}[t]
    \centering
    \caption{Comparison of the cohesive energy (kJ/mol) of selected molecular
crystals at the MP2 level of theory, including local MP2 (LMP2) and the hybrid 
QM/MM many-body interaction (HMBI) model~\cite{wen_accurate_2011}.
The MBE results from Ref.~\onlinecite{sargent_benchmarking_2023} include the monomer relaxation energy from our own calculations as explained in the main text.
    }
    \begin{threeparttable}
    \begin{tabular*}{\linewidth}{l@{\extracolsep{\fill}}ddd}
    \toprule
                   & \ttext{ammonia} & \ttext{CO$_2$} & \ttext{benzene} \\
    \midrule
    this work (PBE-TS structs.)                             & 38.3 & 27.8 & 76.6 \\
    this work (CSD structs.)                           & 33.6  & 28.6 & 72.8 \\
    MBE(2B)-MP2/CBS~\cite{sargent_benchmarking_2023}  & 33.4 & 28.9 & 72.3 \\
    \hline
    LMP2/CBS~\cite{maschio_periodic_2010}    & 35.6 & 29.8 & \ttext{--}  \\
    HMBI-MP2/CBS~\cite{wen_accurate_2011}  & 39.3 & 29.1 & 61.6  \\
    MP2/CBS~\cite{del_ben_second-order_2012}      & 33.9 & 26.1 & 58.7  \\
    LMP2/p-aug-6-31G(d,p)~\cite{cutini_assessment_2016}     & 34.1 & 22.7 & 57.5  \\
    \bottomrule
    \end{tabular*}
\end{threeparttable}
    \label{tab:mp2compare}
    \end{table}

Before comparing to experiment and assessing the accuracy of spin-component scaling,
we first pause to note that---in contrast to periodic, canonical MP2---the many-body expansion (MBE)
has been one of the most popular formalisms for correlated calculations of molecular
crystals~\cite{sargent_benchmarking_2023,metcalf_range-dependence_2022,hofierka_binding_2021}.
Although convergence of MBE calculations is nontrivial~\cite{xie_assessment_2023,kennedy_communication_2014,yang_ab-initio_2014}, a recent paper by Sargent et al.\ reported carefully converged cohesive energies of most of the molecular crystals in the X23 dataset using MBE with only two-body contributions [MBE(2B)]~\cite{sargent_benchmarking_2023}.
This work used structures taken directly from the Cambridge Structural Database (CSD) and did not calculate the one-body relaxation energy, so we have calculated this quantity ourselves for the same crystal geometries, to facilitate comparison.

In table~\ref{tab:mp2compare}, we present a comparison of the cohesive energies of three commonly studied molecular crystals and find that the agreement between our own periodic calculations and the MBE calculations from Ref.~\onlinecite{sargent_benchmarking_2023} is excellent. 
When we use PBE-TS structures (as we do throughout the rest of this work), the differences are 4.9, 1.1, and 4.4~kJ/mol, for ammonia, carbon dioxide, and benzene, respectively.
When we repeat our calculations with the same CSD structures, the difference is
reduced to 0.2, 0.3, and 0.5~kJ/mol.
However, upon extending this comparison to the rest of the X23 dataset, we find that some of the CSD structures used in Ref.~\onlinecite{sargent_benchmarking_2023} have extremely large molecular relaxation energies, yielding abnormally bad cohesive energies. 
We thus suggest that unoptimized CSD geometries should be used with caution, and postpone a detailed comparison with MBE to future work.
We conclude that kJ/mol agreement is possible with tightly converged MBE or periodic calculations, but that alternative geometries can cause differences of about 5~kJ/mol or more.
To demonstrate the challenge of kJ/mol agreement, in the bottom half of table~\ref{tab:mp2compare},
we also show MP2 cohesive energies from a few other reports in the literature, which commonly differ by 5--15~kJ/mol.

\begin{figure}[t]
    \centering
    \includegraphics[width=8cm]{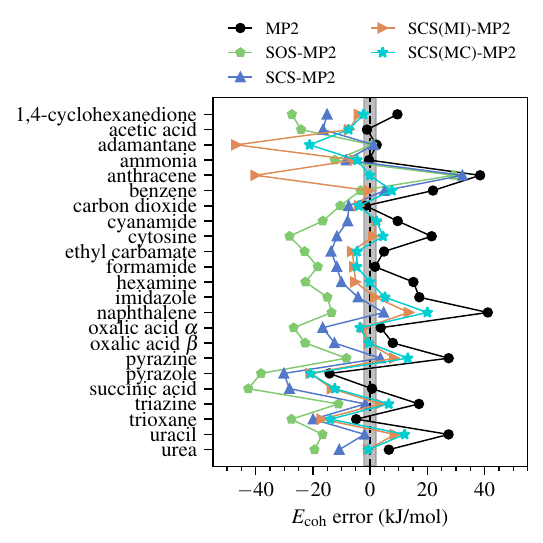}
    \caption{Error in the cohesive energy ($\mathrm{error} = E_\mathrm{coh}^\mathrm{calc}-E_\mathrm{coh}^\mathrm{exp}$) of the molecular crystals in the X23 dataset using MP2 and four modified MP2 models: the scaled-opposite spin (SOS) model\cite{jung_scaled_2004}, the spin component-scaled (SCS) model\cite{grimme_improved_2003}, the SCS-molecular interaction [SCS(MI)] model\cite{distasio_jr_optimized_2007},
    and the SCS-molecular crystal [SCS(MC)] model developed in this work.
    The estimated uncertainty of $\pm 2$ kJ/mol in our calculated numbers is indicated by the grey shaded area.}
    \label{fig:coh_e_diff}
\end{figure}

Having shown an example of our own convergence in figure~\ref{fig:ammonia_ex} and demonstrated good agreement (sub kJ/mol) with similarly converged
MBE results, we now turn to a comparison with experimental cohesive energies.
In figure~\ref{fig:coh_e_diff}, we present the error of the converged MP2 cohesive energy (with PBE-TS geometries) compared to the corrected experimental cohesive energy~\cite{dolgonos_revised_2019} for all 23 molecular crystals.
We see that MP2 overestimates the cohesive energy, which is a well documented behavior of MP2~\cite{distasio_jr_optimized_2007,goldzak_accurate_2022} due to an inaccurate description of dispersion interactions and intermolecular binding energies.
As discussed in the introduction, improved results at the same cost can be obtained using spin-component scaling~\cite{grimme_improved_2003}, wherein the correlation energy is calculated as $E_\mathrm{c}^\mathrm{SCS} = c_{\mathrm{os}}E_{\mathrm{os}}^{(2)} + c_{\mathrm{ss}}E_{\mathrm{ss}}^{(2)}$ and $c_\mathrm{os}, c_\mathrm{ss}$ have been previously optimized on training data.
In figure~\ref{fig:coh_e_diff}, we show results from three different spin scaling prescriptions developed previously:
the scaled-opposite spin (SOS) model\cite{jung_scaled_2004} ($c_{\mathrm{ss}} = 0$, $c_{\mathrm{os}} = 1.3$),
the spin-component-scaled (SCS) model\cite{grimme_improved_2003} ($c_{\mathrm{ss}} = 1/3$, $c_{\mathrm{os}} = 6/5$),
and the SCS-molecular interaction [SCS(MI)] model\cite{distasio_jr_optimized_2007} ($c_{\mathrm{ss}} = 1.29$, $c_{\mathrm{os}} = 0.4$).
We see that all three spin scaling prescriptions correct for the overbinding tendency of MP2.
The magnitude of the corrections displays an overall trend, SOS-MP2 $>$ SCS-MP2 $>$ SCS(MI)-MP2, and is typically too large, resulting in predicted cohesive energies that are, on average, too small compared to experiment.

In table~\ref{tab:error_summ}, we collect performance statistics of MP2 and the spin scaling variants, along with those of DFT results with various dispersion corrections.
We see that SCS(MI)-MP2~\cite{distasio_jr_optimized_2007} delivers the best performance, with a mean absolute error (MAE) of 9.5~kJ/mol and a mean signed error (MSE) of $-6.1$~kJ/mol over the entire dataset.
Remarkably, SCS(MI)-MP2~\cite{distasio_jr_optimized_2007} has a MAE of only 5.4~kJ/mol for the hydrogen-bonded molecular crystals, indicating that purely dispersion bound complexes are the most difficult.
Indeed, with pure MP2, we find the largest errors for anthracene (49.1~kJ/mol) and naphthalene (39.0~kJ/mol), reflecting the challenge of capturing their $\pi$-electron based dispersion interactions with second-order perturbation theory, which is consistent with previous studies of molecular interactions~\cite{shee_regularized_2021,grimme_accurate_2004}.
For dispersion bound crystals, we find that all MP2-based methods have a MAE of about 10--20~kJ/mol.

In contrast, the improved performance for hydrogen-bonded crystals can likely be attributed to the ability of HF to capture some amount of polarization and electrostatics.
Rather remarkably, DFT with the PBE functional and dispersion corrections largely outperforms MP2-based methods.
Specifically, we see that PBE-TS~\cite{tkatchenko_accurate_2009} is slightly worse than MP2-based methods, whereas PBE-D3 and PBE-MBD
are significantly better, exhibiting average errors of only 3-5~kJ/mol.

\begin{table}[b]
    \caption{Error statistics (kJ/mol) of our MP2 results and DFT results for the X23 dataset, including mean absolute energy (MAE) and mean signed error (MSE) compared to experimental values. In the final two columns, we separate performance on crystals dominated by hydrogen bonding (HB) and dispersion (disp) interactions.
}
    \centering
\begin{threeparttable}
    \begin{tabular*}{\linewidth}{l@{\extracolsep{\fill}}dddd}
        \toprule
        Theory      & \ttext{MAE}  & \ttext{MSE}   & \ttext{MAE (HB)} & \ttext{MAE (disp)} \\
        \hline
        MP2         &  12.9 & 11.3  & 5.2      & 18.3       \\
        SCS-MP2     &  11.9 & -7.9  & 14.3     & 10.5       \\
        SOS-MP2     &  19.9 & -17.3 & 23.5     & 15.3       \\
        SCS(MI)-MP2 &  9.5  & -6.1  & 5.4      & 15.0       \\
        SCS(MC)-MP2 &  7.5  & -1.3  & 4.1      & 9.8        \\
        PBE-D3~\cite{reilly_understanding_2013}
                    & 4.6  & 2.9   & 7.1      & 2.6       \\
        PBE-TS~\cite{reilly_understanding_2013}
                    & 13.0 & 12.7  & 10.5     & 16.0      \\
        PBE-MBD~\cite{mortazavi_structure_2018}
                    & 4.5  & 3.1   & 5.0      & 3.9       \\
        \bottomrule
        \end{tabular*}
\end{threeparttable}
\label{tab:error_summ}
\end{table}

In figure~\ref{fig:coh_e_diff}, we see two prominent outliers, adamantane and anthracene, which do not follow the trend seen for the other molecular crystals.
In particular, for these crystals, the correction from both SOS-MP2 and SCS-MP2 nearly vanishes, while that from SCS(MI)-MP2 is abnormally large and negative (i.e., the magnitude of the cohesive energy is significantly underestimated by more than 40~kJ/mol).

This behavior can be traced back to the degree to which a spin-component scaling prescription conserves the magnitude of the MP2 correlation energy (note that energy differences are much smaller than the total correlation energy).
Specifically, we define the ratio
\begin{equation}
\alpha = \frac{E_{\mathrm{SCS}}^{(2)}}{E^{(2)}} = \frac{c_\mathrm{ss} + c_\mathrm{os}\gamma}{1+\gamma}
\end{equation}
where $\gamma = E_{\mathrm{os}} / E_{\mathrm{ss}}$, and $\gamma$ is typically 3--3.5 (see Ref.~\onlinecite{grimme_improved_2003} and figure S6). 
Then a theory that approximately conserves the magnitude of the MP2 correlation energy has $\alpha \approx 1$, which implies
$c_\mathrm{ss} + 3c_\mathrm{os} \approx 4$; this is satisfied for the SOS and SCS models, but not for the SCS(MI) model, which has $\alpha \approx 0.6$.
The SCS(MI) correlation energies are not accurate (far too small), but the method is optimized for energy differences, which can still be accurate.
As detailed in section~V in the SI, in the presence of spin scaling, two terms dominate the correction to the MP2 cohesive energy:
one term proportional to $(\alpha - 1) E^{(2)}_{\mathrm{coh}}$ with $E^{(2)}_{\mathrm{coh}}$ the correlation part of $E_{\mathrm{coh}}$,
and the other term proportional to $(c_{\mathrm{ss}} - c_{\mathrm{os}}) \delta$ with $\delta = \gamma_{\mathrm{mol}} - \gamma_{\mathrm{cell}}$.
Empirically, we observe $\delta \approx 0.1$ for most systems in the X23 set (figure~S6).
For SCS-MP2 and SOS-MP2, with $\alpha \approx 1$, the first term vanishes and the second term dominates the correction, which is negative 
because $c_{\mathrm{ss}} < c_{\mathrm{os}}$, and the cohesive energy is correctly reduced.
By contrast, for SCS(MI)-MP2, where $\alpha \approx 0.6$, both correction terms are nonzero but with opposite sign, so they partially cancel and render the net correction smaller than the other two prescriptions.
However, when $\delta \approx 0$, as is accidentally the case for the two outliers, adamantane and anthracene (figure~S6), the net correction from both SCS-MP2 and SOS-MP2 vanishes, while that from SCS(MI)-MP2 is uncompensated and abnormally large, consistent with our observation from figure~\ref{fig:coh_e_diff}.

Approximate conservation of correlation energy has been widely regarded as a good practice for mixing different energy components, with the most famous example being hybrid DFT~\cite{stephens_abinitio_1994}.
The accidental vanishing of $\delta$ for the two outliers discussed above is thus unfortunate, as the analysis suggests that any spin-component scaling prescription that approximately conserves the MP2 correlation energy is doomed to show little improvement over the bare MP2 cohesive energy.

When $\delta=0$, nonzero corrections can be obtained by a theory that disregards the conservation of the MP2 correlation energy, but such a theory is at risk of uncompensated, abnormally large corrections as seen in SCS(MI)-MP2.

\begin{figure}[t]
    \centering
    \includegraphics[width=5cm]{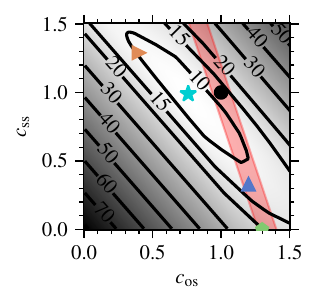}
    \caption{MAE (kJ/mol) for the spin-component scaled MP2 cohesive energy of the X23 dataset.
    Indicated points correspond to the scaling parameters for MP2 (black dot), SOS-MP2 (green pentagon), SCS-MP2 (blue triangle), SCS(MI)-MP2 (orange triangle), and our new SCS(MC)-MP2 (cyan star).
    The red-shaded area highlights the range of spin scaling parameters that approximately conserve the MP2 correlation energy.}
    \label{fig:ssos_mae}
\end{figure}

With all these observations in mind, we assess the extent to which further improvement is possible by a reoptimization of the spin-scaling parameters.
In figure~\ref{fig:ssos_mae}, we plot the cohesive energy MAE as a function of the scaling parameters and identify $c_{\mathrm{ss}} = 0.99$, $c_{\mathrm{os}} = 0.76$ to be optimal for the X23 set; we call this new prescription SCS for molecular crystals [SCS(MC)].
As shown in figure~\ref{fig:coh_e_diff} for all crystals, the cohesive energies predicted by SCS(MC)-MP2 are very similar to those of SCS(MI)-MP2 except for the two outliers, where the significant underestimation by SCS(MI)-MP2 is largely corrected, resulting in an improved MAE (MSE) of $7.5$~kJ/mol ($-1.3$~kJ/mol) as seen in table~\ref{tab:error_summ}. In figure~\ref{fig:ssos_mae}, we also show the region of parameters that conserve the MP2 energy (i.e., those with $\alpha\approx 1$ or with $c_\mathrm{ss} + 3c_\mathrm{os} \approx 4$). We see that the new SCS(MC) parameters do not violate this constraint nearly as much as those of SCS(MI), and empirically we find $\alpha \approx 0.8$. This compromise allows for reasonable correlation energy conservation while maintaining the freedom to correct the MP2 cohesive energy even when $\delta =0$.

In summary, we have presented cohesive energies of 23 molecular crystals at the MP2 level of theory, with careful attention paid to basis set and finite-size errors.
For our chosen geometries, we believe our results are converged to better than 2~kJ/mol.
We hope that these results will serve as useful benchmarks in future applications of wavefunction methods for molecular crystals and other solids.
Importantly, we have emphasized that the cohesive energy of different but reasonable geometries can differ by 5~kJ/mol or more, which presents a challenge for precise comparisons between calculated or experimental values.

With regards to performance, we have demonstrated the well-known trend of MP2 to overbind, resulting in an overestimation of the cohesive energy by about 10--20~kJ/mol on average.
Separate scaling of spin components approximately halves this error, making predictions with 5--10~kJ/mol accuracy possible.
Although we proposed a new spin scaling prescription, we note that optimization against experimental values is imperfect, for reasons discussed throughout the text, and the degree to which this new model is transferable to other problems is unknown.
Although we have demonstrated that kJ/mol accuracy cannot be reliably obtained within the family of spin-component scaled MP2 methods, we anticipate applications of regularized MP2~\cite{shee_regularized_2021} or double-hybrid DFT~\cite{sharkas_double-hybrid_2014,stein_double-hybrid_2020,wang_doubly_2021} as possible avenues towards kJ/mol accuracy without the cost of coupled-cluster theory~\cite{yang_ab-initio_2014,sherrill_2023}. 

\section*{Associated Content}
\subsection*{Supporting Information}
The Supporting Information is available free of charge at [publisher inserts link].

\vspace{1em}

Cohesive energies for all 23 crystals at all levels of theory discussed, basis set and thermodynamic limit convergence testing, and discussion on the spin scaling correction to the cohesive energy.

\section*{Notes}
The authors declare no competing financial interest.

\section*{Acknowledgements}
We thank Caroline Sargent and Prof.\ David Sherrill for useful discussions and for sharing
MP2/CBS data from Ref.~\onlinecite{sargent_benchmarking_2023}.
This work was supported by the National Science Foundation under
Grant Nos.~CHE-1848369 and OAC-1931321.
We acknowledge computing resources from the Flatiron Institute Scientific Computing Center
and Columbia University's Shared Research
Computing Facility project, which is supported by NIH Research Facility
Improvement Grant 1G20RR030893-01, and associated funds from the New York State
Empire State Development, Division of Science Technology and Innovation
(NYSTAR) Contract C090171, both awarded April 15, 2010.

\end{document}


\title{Supporting Information for:
Can spin-component scaled MP2 achieve kJ/mol accuracy for cohesive energies of molecular crystals?
}

\author{Yu Hsuan Liang}
\affiliation{Department of Chemistry, Columbia University, New York, NY 10027 USA}
\author{Hong-Zhou Ye}
\email{hzyechem@gmail.com}
\affiliation{Department of Chemistry, Columbia University, New York, NY 10027 USA}
\author{Timothy C. Berkelbach}
\email{t.berkelbach@columbia.edu}
\affiliation{Department of Chemistry, Columbia University, New York, NY 10027 USA}

\maketitle
Detailed data can be found in the \href{https://github.com/welltemperedpaprika/x23_mp2}{GitHub repository}.

\section{$k$-point sampling mesh pairs for TDL extrapolation}
\begin{figure}[h!]
    \centering
    \includegraphics[scale=0.9]{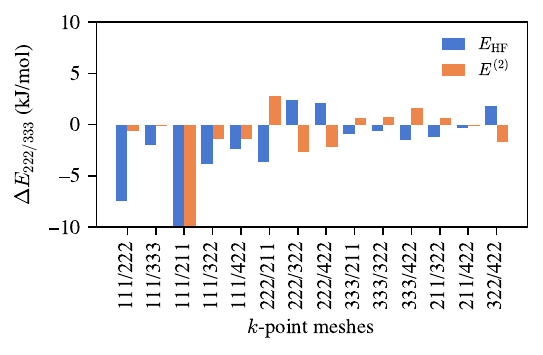}
    \caption{Extrapolation errors from ``viable'' $k$-point mesh pairs for ethyl carbamate crystal, whose lattice constants are $a =  5.009~\text{\AA}$, $b =  6.954~\text{\AA}$, and $c =  7.418~\text{\AA}$. From this data, we choose 211/322 as the most efficient $k$-point mesh pair for production calculations.
    }
    \label{fig:kptsamp}
\end{figure}

The most computationally efficient pair of $k$-point sampling meshes for extrapolation is obtained by screening a range of ``viable'' $k$-point meshes. 
We define ``viable'' meshes as those for which the number of $k$-points along each axis is roughly inversely proportional to the corresponding lattice constants of the unit cell. 
All such exploratory calculations are done with a minimal basis set, and the most efficient choice of $k$-point mesh pairs for production calculations is determined from the smallest $k$-point mesh pair with an extrapolation error smaller than 2~kJ/mol with respect to a dense $k$-point mesh pair.
An example for ethyl carbamate is shown in figure~\ref{fig:kptsamp}. 
\section{Composite Correction}

\begin{figure}[h!]
    \centering
    \includegraphics[scale=0.8]{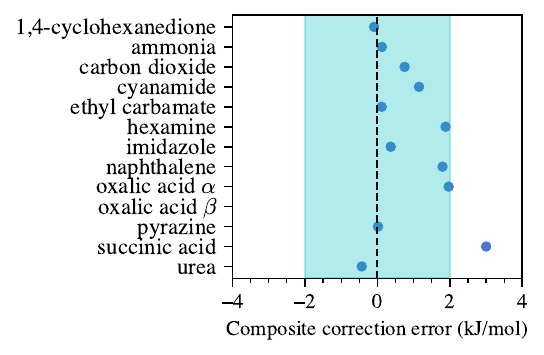}
    \caption{Estimated composite correction errors in the cohesive energy from 13 molecular crystals where the large $N_{k,2}$/QZ calculation is possible. 
    The mean error is 1.18~kJ/mol.
}
    \label{fig:comp_corr_all}
\end{figure}

As discussed in the main text, when the calculation at the larger $k$-point mesh with $N_{k,2}$ points is not possible in the QZ basis, we apply the composite correction in equation~(5). 
The error of this approximation can be assessed on molecular crystals for which the large calculation is possible (13 crystals in total). 
In the main text figure 2, we showed an example for ammonia including the estimated TDL/CBS cohesive energy using the composite corrected $N_{k,2}$/QZ value. 
The error in the cohesive energy due to this composite correction is shown in figure~\ref{fig:comp_corr_all} for all 13 crystals for which it is accessible.

\newpage

\section{Basis Set Convergence}

\begin{figure}[h!]
    \centering
    \includegraphics[scale=0.9]{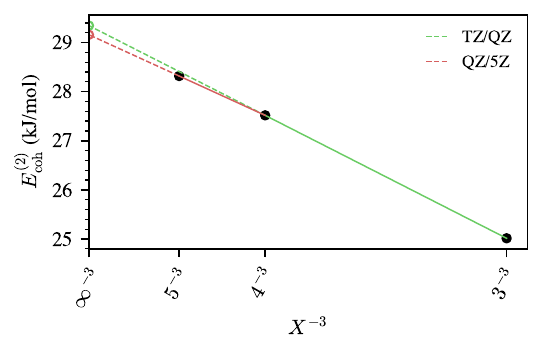}
    \caption{Basis set convergence for the ammonia crystal at $k=222$. The hollow circles represent extrapolated values from a given basis set pair. Similar behavior is seen for other $k$-point meshes.}
    \label{fig:basis}
\end{figure}

Results reported in the main text were obtained using TZ/QZ extrapolation.
In figure~\ref{fig:basis}, we show basis set convergence for the ammonia crystal at two different $k$-point mesh sizes.
At both mesh sizes, we see that TZ/QZ extrapolation has an error of less than 0.5~kJ/mol compared to QZ/5Z extrapolation.
We performed similar analysis (i.e., using small $k$-point meshes) for all 23
molecular crystals and concluded that TZ/QZ extrapolation has a mean error of
0.9~kJ/mol compared to QZ/5Z extrapolation, justifying our use of the former in
the main text.

\widetext

\section{MP2, SCS-MP2, SOS-MP2, SCS(MI)-MP2 Results}

Numerical values of the cohesive energies of all 23 molecular crystals at all discussed levels
of theory (plus experiment) are given in table~\ref{tab:ecoh}.

\begin{table*}[!h]
    \centering
    \label{tab:ecoh}
    \caption{Cohesive energies (kJ/mol) at TDL and CBS limit for all X23 molecular crystals.}
    \begin{tabular}{lrrrrrrr}
        \toprule
        {} &     HF &     MP2 &  SCS-MP2 &  SOS-MP2 &  SCS(MI) &  SCS(X23) &    Exp \\
        molecule             &        &         &          &          &         &          &        \\
        \midrule
        1,4-cyclohexanedione &     1.45 &   99.54 &    74.96 &    62.67 &   86.07 &    87.73 &   90.0 \\
        acetic acid          &    16.64 &   72.50 &    57.14 &    49.47 &   66.22 &    66.13 &   73.6 \\
        adamantane           & $-$52.88 &   74.10 &    73.05 &    72.52 &   25.07 &    50.64 &   71.8 \\
        ammonia              &     4.76 &   38.34 &    30.33 &    26.33 &   33.31 &    34.19 &   38.7 \\
        anthracene           & $-$66.15 &  148.84 &   142.64 &   139.54 &   70.37 &   110.30 &  110.4 \\
        benzene              & $-$24.23 &   76.57 &    59.68 &    51.23 &   54.13 &    62.21 &   54.8 \\
        carbon dioxide       &     6.88 &   27.81 &    21.92 &    18.98 &   25.59 &    25.46 &   29.4 \\
        cyanamide            &    26.94 &   91.13 &    73.67 &    64.94 &   83.74 &    83.77 &   81.5 \\
        cytosine             &    47.20 &  185.02 &   151.89 &   135.33 &  164.65 &   168.05 &  163.5 \\
        ethyl carbamate      &    16.61 &   93.05 &    74.56 &    65.32 &   81.87 &    83.67 &   88.2 \\
        formamide            &    29.78 &   82.82 &    69.48 &    62.82 &   75.58 &    76.44 &   81.1 \\
        hexamine             & $-$17.73 &   99.23 &    74.06 &    61.48 &   78.93 &    84.05 &   84.1 \\
        imidazole            &    12.87 &  107.61 &    86.19 &    75.48 &   92.21 &    95.58 &   90.4 \\
        naphthalene          & $-$42.26 &  122.42 &    86.03 &    67.83 &   94.80 &   101.29 &   81.3 \\
        oxalic acid $\alpha$ &    35.07 &  102.51 &    82.23 &    72.09 &   96.73 &    95.29 &   98.8 \\
        oxalic acid $\beta$  &    30.39 &  104.73 &    84.33 &    74.12 &   96.35 &    96.25 &   96.8 \\
        pyrazine             & $-$18.44 &   91.75 &    67.92 &    56.01 &   72.74 &    77.48 &   64.3 \\
        pyrazole             &     6.27 &   64.61 &    48.67 &    40.69 &   57.97 &    57.94 &   78.8 \\
        succinic acid        &    20.23 &  130.77 &   101.94 &    87.52 &  116.76 &   117.76 &  130.1 \\
        triazine             &  $-$3.33 &   79.67 &    60.91 &    51.53 &   66.19 &    69.14 &   62.6 \\
        trioxane             & $-$10.10 &   59.73 &    44.66 &    37.12 &   47.65 &    50.68 &   64.6 \\
        uracil               &    39.69 &  163.62 &   134.26 &   119.58 &  144.86 &   148.24 &  136.2 \\
        urea                 &    45.19 &  108.66 &    91.33 &    82.66 &  101.42 &   101.40 &  102.1 \\
        \bottomrule
        \end{tabular}        
\end{table*}

\newpage

\section{Effect of spin scaling on cohesive energy}

Let $\gamma = E_{\mathrm{os}}^{(2)} / E_{\mathrm{ss}}^{(2)}$.
The MP2 correlation energy can be written as
\begin{equation}
    E^{(2)}
        = E_{\mathrm{ss}}^{(2)} + E_{\mathrm{os}}^{(2)}
        = ( 1 + \gamma ) E_{\mathrm{ss}}^{(2)}
\end{equation}
Let $X$ denote spin scaling prescription specified by coefficients $(c_{\mathrm{ss}}^{X}, c_{\mathrm{os}}^{X})$, which are listed in table~\ref{tab:css_cos} for the three prescriptions used in this work.

\begin{table}[!h]
    \centering
    \caption{Spin scaling coefficients for different spin scaling prescriptions.}
    \label{tab:css_cos}
    \begin{tabular}{lll}
        \toprule
        {}  & $c_{\mathrm{ss}}$ & $c_{\mathrm{os}}$ \\
        \midrule
        SOS & $0$ & $1.3$   \\
        SCS & $0.333$ & $1.2$   \\
        SCS(MI) & $1.29$ & $0.4$   \\
        SCS(MC) & $0.99$ & $0.76$   \\
        \bottomrule
    \end{tabular}
\end{table}

The correlation energy after spin scaling can be written as
\begin{equation}
\begin{split}
    E^{(2),X}
        = c_{\mathrm{ss}}^{X} E_{\mathrm{ss}}^{(2)} +
        c_{\mathrm{os}}^{X} E_{\mathrm{os}}^{(2)}
        = (c_{\mathrm{ss}}^{X} + c_{\mathrm{os}}^{X} \gamma) E_{\mathrm{ss}}^{(2)}
        = \underbrace{
            \frac{c_{\mathrm{ss}}^{X} + c_{\mathrm{os}}^{X} \gamma}{1 + \gamma}
        }_{\alpha^{X}(\gamma)}
        E^{(2)}
        = \alpha^{X}(\gamma) E^{(2)}.
\end{split}
\end{equation}
The spin scaling correction to the cohesive energy reads
\begin{equation}
    \Delta E_{\mathrm{coh}}^{X}
        = E^{X}_{\mathrm{coh}} - E_{\mathrm{coh}}
        \approx \left[
            \alpha^{X}(\gamma_{\mathrm{mol}}) E_{\mathrm{mol}}^{(2)} -
            \alpha^{X}(\gamma_{\mathrm{cell}}) \frac{E_{\mathrm{cell}}^{(2)}}{N_{\mathrm{mol}}}
        \right] -
        \left[
            E_{\mathrm{mol}}^{(2)} - \frac{E_{\mathrm{cell}}^{(2)}}{N_{\mathrm{mol}}}
        \right]
\end{equation}
where we have discarded the relaxation energy whose contribution to $\Delta E_{\mathrm{coh}}^{X}$ is small.

\begin{figure}[!b]
    \centering
    \includegraphics[scale=0.8]{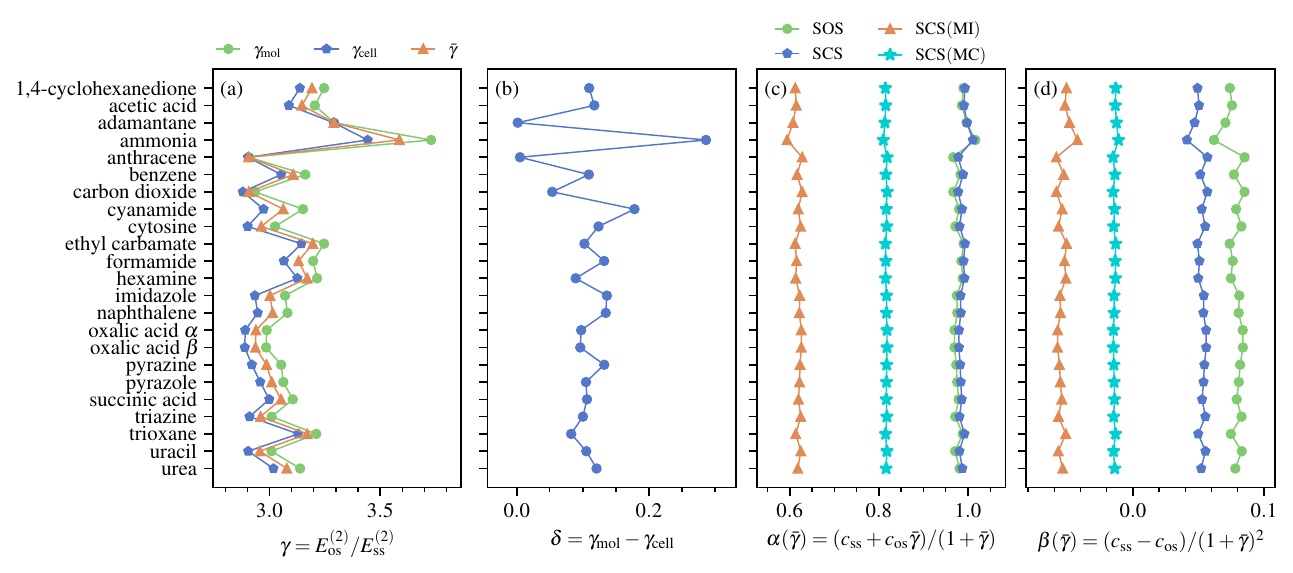}
    \caption{Plot of parameters $\gamma$ (a), $\delta$ (b), $\alpha(\bar{\gamma})$ (c), and $\beta(\bar{\gamma})$ (d) discussed in the text for all systems in the X23 set.}
    \label{fig:gdab_all}
\end{figure}

\begin{figure}[!b]
    \centering
    \includegraphics[scale=0.8]{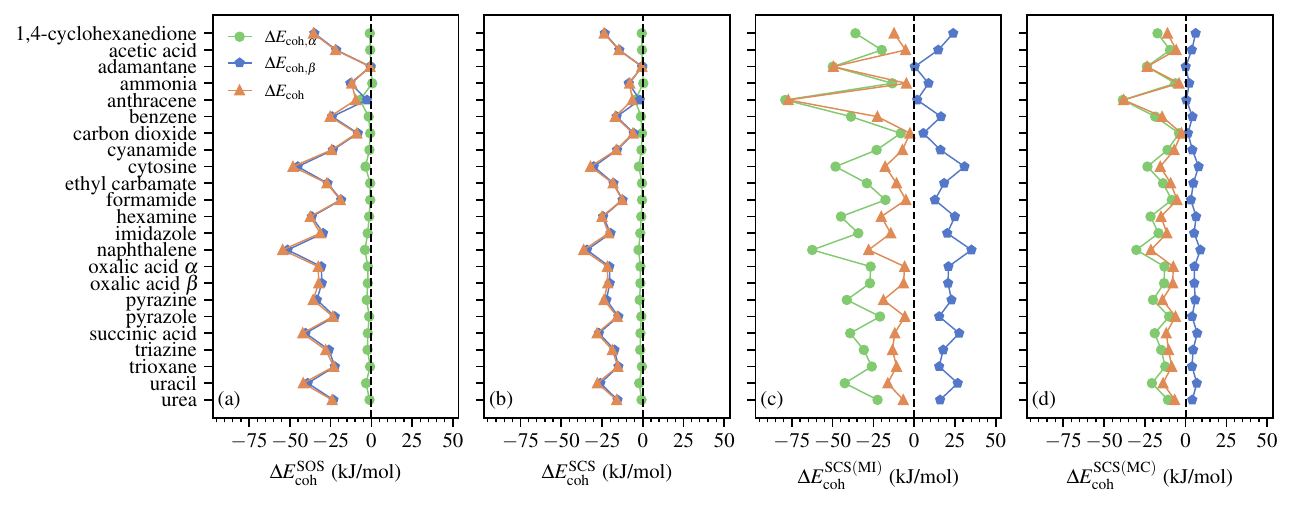}
    \caption{Plot of the components $\Delta E_{\mathrm{coh},\alpha}^{X}$ and $\Delta E_{\mathrm{coh},\beta}^{X}$ of the spin scaling correction $\Delta E_{\mathrm{coh}}^{X}$ defined in equation~\ref{eq:ecoh_decomp} for (a) SOS, (b) SCS, (c) SCS(MI), and (d) SCS(MC) for all systems in the X23 set.}
    \label{fig:ecoh_decomp_all}
\end{figure}

The values of $\gamma_{\mathrm{mol}}$ and $\gamma_{\mathrm{cell}}$ for the entire X23 set are plotted in figure~\ref{fig:gdab_all}a.
We see that typically $\gamma_{\mathrm{mol}} \approx \gamma_{\mathrm{cell}} \approx 3$, motivating the introduction of the following two parameters
\begin{equation}
    \bar{\gamma}
        = \frac{\gamma_{\mathrm{mol}} + \gamma_{\mathrm{cell}}}{2},
    \qquad{}
    \delta
        = \gamma_{\mathrm{mol}} - \gamma_{\mathrm{cell}}
\end{equation}
The parameter $\delta$ is small but positive for the entire X23 set (figure~\ref{fig:gdab_all}b), which justifies a linear expansion of $\Delta E_{\mathrm{coh}}^{X}$ around $\bar{\gamma}$
\begin{equation}    \label{eq:ecoh_decomp}
    \Delta E_{\mathrm{coh}}^{X}
        \approx \underbrace{
            \left[ \alpha^{X}(\bar{\gamma}) - 1 \right] E^{(2)}_{\mathrm{coh}}
        }_{\Delta E_{\mathrm{coh},\alpha}^{X}}
        + \underbrace{
            \beta^{X}(\bar{\gamma}) \frac{E^{(2)}_{\mathrm{cell}}}{N_{\mathrm{mol}}} \delta
        }_{\Delta E_{\mathrm{coh},\beta}^{X}}
        + O(\delta^2)
        \approx \Delta E_{\mathrm{coh},\alpha}^{X} + \Delta E_{\mathrm{coh},\beta}^{X}
\end{equation}
where
\begin{equation}
    \beta^{X}(\gamma)
        = \frac{
            c_{\mathrm{ss}}^{X} - c_{\mathrm{os}}^{X}
        }{
            (1 + \gamma)^2
        }.
\end{equation}
The first term $\Delta E_{\mathrm{coh},\alpha}^{X}$ thus reflects how much the spin scaled correlation energy deviates from the bare MP2, while the second term is proportional to $(c_{\mathrm{ss}}^{X} - c_{\mathrm{os}}^{X}) \delta$.
From figure~\ref{fig:gdab_all}(c) we see that $\alpha^{\mathrm{SOS}} \approx \alpha^{\mathrm{SCS}} \approx 1$, consistent with the two methods approximately conserving the MP2 correlation energy, while $\alpha^{\mathrm{SCS(MI)}} \approx 0.6$.
As a result, $\Delta E_{\mathrm{coh},\alpha}^{X}$ vanishes for both SOS and SCS-MP2, while being a large negative value for SCS(MI)-MP2.
The second term, on the other hand follows a trend
\begin{equation}
    \Delta E_{\mathrm{coh},\beta}^{\mathrm{SOS}} >
        \Delta E_{\mathrm{coh},\beta}^{\mathrm{SCS}}
        \approx -\Delta E_{\mathrm{coh},\beta}^{\mathrm{SCS(MI)}}
        > 0
\end{equation}
where the sign change for SCS(MI)-MP2 is due to the flipped spin scaling coefficients (table~\ref{tab:css_cos}).
Both the two components, $\Delta E_{\mathrm{coh},\alpha}^{X}$ and $\Delta E_{\mathrm{coh},\beta}^{X}$, and the total spin scaling correction $\Delta E_{\mathrm{coh}}^{X}$ are plotted in figure~\ref{fig:ecoh_decomp_all}.
As already analyzed in the main text, for most systems where $\delta \approx 0.1$, the cancellation of the two terms in SCS(MI)-MP2 renders its correction the smallest among the three, while for the two outliers, adamantane and anthracene, where $\delta$ vanishes (accidentally), the correction from SOS and SCS-MP2 vanishes accordingly while that from SCS(MI)-MP2 is uncompensated and abnormally large.

